# The challenges of statistical patterns of language: the case of Menzerath's law in genomes.

Ramon Ferrer-i-Cancho[1,*], Núria Forns[2], Antoni Hernández-Fernández[1,3], Gemma Bel-Enguix[4] & Jaume Baixeries[1]

[1]Complexity and Quantitative Linguistics Lab. Departament de Llenguatges i Sistemes Informàtics. TALP Research Center/LARCA. Universitat Politècnica de Catalunya. Barcelona (Catalonia), Spain.

[2]Departament de Microbiologia, Facultat de Biologia, Universitat de Barcelona. Barcelona (Catalonia), Spain.

[3]Departament de Lingüística General. Universitat de Barcelona. Barcelona (Catalonia), Spain.

[4]Laboratoire d'Informatique Fondamentale, University Aix-Marseille & CNRS. Marseille, France.

ABSTRACT

The importance of statistical patterns of language has been debated over decades. Although Zipf's law is perhaps the most popular case, recently, Menzerath's law has begun to be involved. Menzerath's law manifests in language, music and genomes as a tendency of the mean size of the parts to decrease as the number of parts increases in many situations. This statistical regularity emerges also in the context of genomes, for instance, as a tendency of species with more chromosomes to have a smaller mean chromosome size. It has been argued that the instantiation of this law in genomes is not indicative of any parallel between language and genomes because (a) the law is inevitable and (b) non-coding DNA dominates genomes. Here mathematical, statistical and conceptual challenges of these criticisms are discussed. Two major conclusions are drawn: the law is not inevitable and languages also have a correlate of non-coding DNA. However, the wide range of manifestations of the law in and outside genomes suggests that the striking similarities between non-coding DNA and certain linguistics units could be anecdotal for understanding the recurrence of that statistical law.

Kewords: statistical laws, language, genomes, music, non-coding DNA, Menzerath's law.

* Address correspondence to: Ramon Ferrer-i-Cancho, Complexity and Quantitative Linguistics Lab, Departament de Llenguatges i Sistemes Informàtics, TALP Research Center, Universitat Politècnica de Catalunya, Campus Nord, Edifici Omega, Jordi Girona Salgado 1-3, 08034 Barcelona (Catalonia), Spain. E-mail: rferrericancho@lsi.upc.edu. Phone: +34 934137870. Fax: +34 934137787

# 1. Introduction

Attempts to demonstrate that statistical patterns of language have a trivial explanation have a long history that goes back at least to the research by G. A. Miller and collaborators questioning the relevance of Zipf's law for word frequencies around 1960 [1-3]. Zipf's law states that the curve that relates the frequency of a word *f* and its rank *r* (the most frequent word having rank 1, the 2$^{nd}$ most frequent word having rank 2 and so on) should follow $f \sim r^{-\alpha}$ [4]. Miller argued that if monkeys were chained *"to typewriters until they had produced some very long and random sequence of characters"* one would find *"exactly the same "Zipf curves" for the monkeys as for the human authors"* [3]. Under his view, Zipf's law would be an inevitable consequence of the fact that words are made of units, e.g., letters or phonemes. The typewriter argument has been revived many times since then [5-7]. However, rigorous analyses indicate that the curves do not really look the same and the parameters of this random typing model giving a good fit to real word frequencies are not forthcoming [8,9]. Here, we review a recent claim that the finding of another statistical pattern of language, Menzerath's law, is also inevitable [10].

P. Menzerath hypothesized that "the greater the whole, the smaller its constituents" ("*Je größer das Ganze, desto kleiner die Teile*") in the context of language [11] (pp. 101). Converging research in music and genomes [12-15] suggests that Menzerath's law is a general law of natural and human-made systems. In this article, we leave the term Menzerath-Altmann law for referring to the exact mathematical dependency that has been proposed by the quantitative linguistics tradition for the relationship between *x*, the size of the whole (in parts) and *y*, the mean size of the parts, i.e. [16],

$$y = ax^b e^{cx}, \qquad (1)$$

where *a*, *b*, and *c* are the parameters of Menzerath-Altmann law.

In the pioneering research by Wilde & Schwibbe [13] and later work [14,17], Menzerath's law emerged as a negative correlation between $L_c$ and $L_g$, where $L_c$ is the mean chromosome length (the size of the constituents) and $L_g$ is the chromosome number (the size of the construct measured in constituents). More recently, the law has been found in the dependency between mean exon size (the size of the constituents) and the number of exons of human genes (the size of the construct).[15]

However, it has been argued that this negative correlation is trivial [10]: the definition of $L_c$ as a mean, i.e. $L_c = G/L_g$ leads (according to Ref. [10]) unavoidably to $L_c \sim L_g^b$ with *b* = -1, which is supported by fact that mammals and plants give values of *b* that are very close to *b* = -1 (*b* = -1.04 for mammals and *b* = -1.07 for plants [10]). In the present article, ~ is used to indicate proportionality. Furthermore, it has also been argued that a proper connection between human language and genomes cannot be established a priori using genomes as wholes and chromosomes as parts, due to the fluid nature of chromosomal arrangements and the vast dominance of non-coding DNA, which has no parallel in language[10].

Revising those arguments is critical for musicology, quantitative linguistics and genomics. If they were correct, the relationship between the mean size of the constituents (*y*) and the number of constituents (*x*) which have been the subject of many studies [12,15,16,18] would be a trivial consequence of the definition of the size of the constituents as a mean. Following Miller's argument, producing Menzerath's law would be as easy as producing Zipf's law by monkeys chained to a typewriter. More precisely, the inevitability of $L_c \sim 1/L_g$ [10] predicts that Menzerath-Altmann law must always be Eq. (1) with *b* = -1 and *c* = 0 when defining the size of the parts as a mean. If such inevitability is correct, exponents deviating significantly from *b* = -1 should be the exception, not the rule in language, music and genomes.

Here we address the challenge of Menzerath's law in genomes[13-15] and beyond[12,16,18] by reviewing Solé's criticisms[10]: his mathematical and statistical arguments, essentially the inevitability of $L_c \sim 1/L_g$ (Section 2), as well as his conceptual arguments, mainly the mismatch between human language and genomes (Section 3). Finally, we will discuss some general questions that are crucial for understanding the recurrence of Menzerath's law (Section 4).

## 2. The mathematical and statistical debate.

2.1. Mixing angiosperm and gymnosperm plants.

Solé does not distinguish between angiosperm and gymnosperm plants [10]. However, our analyses have been revealing important differences between them: (1) concerning the relationship between $L_g$ and $L_c$, Menzerath's law is only found in angiosperms [14], (2) *G* tends to increase as $L_g$ increases in gymnosperms but *G* increases as $L_g$ decreases in angiosperms [19] and, (3) the fit of $L_c \sim L_g^b$ yields *b* = -0.95 ± 0.05 for angiosperms and *b* = -0.3 ± 0.2 for gymnosperms [17], the latter being statistically inconsistent with *b* = -1 as Solé predicts [10]. As his division of plants differs from that of Ferrer-i-Cancho & Forns [14] and gymnosperms do not follow Menzerath's law, we proceed assuming that his notion of plant is equivalent or can be reduced to angiosperms.

2.2. $L_c = G/L_g$ does not imply $L_c \sim 1/L_g$.

It has been argued that the definition of $L_c$ as $G/L_g$ unavoidably leads to an inverse proportionality dependency between $L_g$ and $L_c$, i.e. $L_c \sim 1/L_g$ [10]. This can be refuted in two ways: empirically and mathematically.

2.2.1. Empirical refutation

$L_c \sim 1/L_g$ is not inevitable because

- Amphibians exhibit a positive correlation between $L_c$ and $L_g$ that is incompatible with $L_c \sim 1/L_g$ [14].
- Menzerath's law (a significant negative correlation between $L_c$ and $L_g$) was not found for gymnosperm plants and ray-finned fishes.
- Many empirical studies of Menzerath-Altmann law compute the size of the parts as an average as Ferrer-i-Cancho & Forns did [14] but the fit of Eq. 1 gives parameters that deviate from b ≈ -1 (see Table 1 for a summary of research).

- *b* = -0.6 is reported for ants in the pioneering work by Wilde & Schwibbe [13] that is cited by Ferrer-i-Cancho & Forns (2009).
- Solé reports estimates of *b* only for mammals and plants (according to his analysis *b* = -1.04 and *b* = -1.07, respectively) [10], whereas Ferrer-i-Cancho & Forns [14], considered a total of eleven major groups [14] (see also [19]). Thus, nine groups have not been considered. |*b*+1| is a measure of the deviation from his prediction, i.e. $L_c \sim 1/L_g$. |*b*+1| = 0 means a perfect matching with his prediction. |*b*+1| indicates that mammals and angiosperm plants are among the three groups with the smallest value of |*b*+1| (Table 2).
- A careful statistical analysis reveals that *b* deviates significantly from *b* = -1 in fungi, gymnosperm plants, insects, reptiles, jawless fishes, ray-finned fishes and amphibians, groups for which Solé reports no result [10]. Furthermore, the parameter *b* of $L_c \sim L_g^b$ contributes significantly to improve the quality of the fit with regard to that of $L_c \sim 1/L_g$ for the same groups [17]. Put differently, if *b* is let free, then the error of the model is reduced significantly for these groups with regard to keeping it equal to -1.
- In a recent study of Menzerath-Altmann law in genomes at the gene-exon level, the relationship between the mean exon size in bases and the number of exons of a human gene yields *b* ≈ -0.5 [15].

2.2.2. Mathematical refutation.

When *G* is a constant function (*G* ~ 1), we have that $L_g \sim L_g^b$ with *b* = -1 as it is argued by Solé [10]. Yet, if *G* is not constant, then *b* = -1 is not necessarily expected: (1) the exponent may change (e.g., if $G \sim L_g^{-2}$ then *b* = 3) and (2) the power-law $L_c \sim L_g^b$ could be lost (e.g., if $G \sim L_g e^{-L_g}$ then we would have $L_c \sim e^{-L_g}$).

A mathematical analysis indicates that $L_c \sim 1/L_g$ needs that *G* and $L_g$ are uncorrelated [19]. Therefore, $L_c \sim 1/L_g$ is rejected if *G* and $L_g$ are correlated. The empirical evidence for such correlation is the following: (1) *G* tends to increase as $L_g$ increases in gymnosperm plants and animals while *G* tends to decrease as $L_g$ decreases in angiosperm plants [19] and (2), from the major taxonomic groups considered by [14], only birds and cartilaginous fishes show no significant correlation between *G* and $L_g$ [19].

2.3. The dependency between $L_c$ and $L_g$.

So far we have been discussing the fit of $L_g \sim L_g^b$ with Solé's prediction of *b*=-1 to genomes. But we have never argued that the instantiation of Menzerath-Altmann law in genomes (recall Eq. 1):

$$L_c = aL_g^b e^{cL_g}, \qquad (2)$$

(with the possibility of *b* = -1 and/or *c* = 0, following Solé's arguments) is the best, or simply the most suitable for modeling the actual relationship between $L_c$ and $L_g$ in genomes. When preparing our original article [14], we were already aware of the challenge of designing biologically realistic equations and evaluating the goodness of their fit rigorously.

Therefore, we decided to use a simple correlation analysis between $L_c$ and $L_g$ to stay neutral about the actual dependency. While our original approach was non-parametric (based on a Spearman rank correlation test), Solé followed the parametric track with the assumption that genomes follow $L_c \sim L_g^b$ [10]. Our approach to test Menzerath's law [14] and our approach to reject $L_c \sim 1/L_g$ are both non-parametric [19]. In sum, our analysis requires fewer assumptions than his. However, we have had to follow a parametric approach in one of the branches of our genome research to show that even when strong assumptions are made about the actual dependency, his arguments do not stand, even for mammals and plants [17].

## 3. The conceptual debate.

3.1. The unsupported fluid nature of chromosomal rearrangements.

Solé states that "*the fluid nature of chromosomal rearrangements through time rules against any special multiscale link between genome-level and chromosome-level patterns*" [10]. If the mathematical interpretation of this statement is that the genome and the chromosome level are statistically independent, then a large amount of research indicates that $G$ and $L_g$ are not independent in real genomes and that independence is in conflict with chromosome well-formedness (see [20] and references therein).

3.2. Languages also have "*dark matter*".

Solé argues that the dominance of non-coding DNA (what he also calls "*information-lacking DNA*", "*dark matter*" or "*junk DNA*"), should prevent us from using large-scale structures such as genomes as meaningful information-related units [10]. However, the view of non-coding DNA as "*dark matter*" or "*junk*" in a strict sense is outdated from the point of view of molecular biology [21-24]. Some researchers have suggested that "*there is in fact much less, if any, "junk" in the genomes of the higher organisms than has previously been supposed*" [25].

Linguistic sequences and genomes are not so radically different concerning real or apparent "*junk*", "*dark matter*" or "*information-lacking DNA*". In general, words are classified into content, e.g., verbs, nouns, and function words, e.g., prepositions, conjunctions. While content words are said to have lexical meaning, function words are said to have grammatical meaning [26], i.e. function words lack lexical meaning [27] (pp. 55). For this reason they are called "*empty words*" by certain scholars [26]. Similarly, non-coding DNA is *empty*, in the sense that it does not code for specific proteins. The term "junk words" has also been used for referring to function words and particles in language sciences [28]. However, the closest analogy for the term "junk" in human language are the so-called filler words such as "um", "oh", "well" [29].

Function words such as prepositions and conjunctions have an inherently relational meaning [30] and they are very important nodes in word networks: they are hubs or "authorities" in a network theory sense [31,32]. The logic structure of the sentence "Mary bought an apartment *in spite of* the economic crisis" is radically different from that of "Mary bought an apartment *thanks to* the economic crisis". The conjunctions "in spite of" and "thanks to" regulate the relationship between "Mary bought an apartment" and "the economic crisis" in the sentences above. In sum, *lexical meaning* and *protein coding* appear to be parallel terms, respectively, from the linguistic and genetic world. The same applies to *grammatical meaning* and *regulation*, the latter being a function served by non-coding DNA[22,24].

If we consider linguistic units with grammatical function as equivalent to non-coding DNA, then not only function words or particles parallel non-coding DNA, but also bound morphemes (e.g., the *–ed* ending of *walked*), as they also contain grammatical meaning. As linguistic sequences at many levels contain a mixture of elements with lexical and grammatical meaning (e.g., lexemes and bound morphemes in words), a DNA sequence may be a combination of coding and non-coding parts (e.g., exons and introns in genes). Words, phrases, clauses, sentences..., i.e. units on which Menzerath's law has been reported ([33] and references therein), are "polluted" to some extent by "dark matter".

The statistics of the amount of function and content words provides us with an estimate of the amount of "dark matter" in language. Table 3 indicates that the proportion of a parallel of non-coding DNA in an English conversation is about 59%, which includes function and filler words, while it is about 37% in a news report. Therefore, languages also have a large proportion of elements reminiscent of non-coding DNA. But the true proportion of "non-coding" elements in languages could be higher if the grammatical morphemes that are attached to lexemes were included in the counts.

Interestingly, the evolution of the view of "fillers" in linguistics parallels the evolution of the view of non-coding regions in molecular biology. Progress in linguistic research indicates that "fillers" are more than mere "fillers" while progress in genomics indicates that "junk" DNA is more than mere "junk". As for linguistics, the understanding of filler words in linguistics has evolved from the term filler [34], as their meaning and their role in the sentence was gradually recognized, to particular kinds of discourse related particles or cue words ([35] and references therein). At present, the consensus is that "words" originally called fillers "*have no apparent grammatical relation to the sentences in which they appear*", and "*contrary to what prescriptivists' accusations, they do have a meaning, in that they seem to convey something about the speaker's relation to what is asserted in the sentence*" [35]. The evolution of the view of other function words has also evolved similarly: function words believed to be empty contain indeed meaning [35,36]. As for molecular biology, the field is moving from the view of non-coding DNA as "junk" to that of functionally relevant material [21-24]. The view of repetitive segments in DNA sequences as mere "fillers" is being abandoned in molecular biology [37]. In both biology and linguistics, "dark matter" is becoming meaningful or functional matter, thanks to progress in core molecular biology and linguistics.

3.3. Misunderstanding of a metaphor

Solé's focus on non-coding DNA as an obstacle for a proper connection between human language and genomes [10] shows that he has misunderstood the "*metaphor that genomes are words and chromosomes are syllables*" (abstract of [14]).

Patterning consistent with Menzerath's law is found at many linguistic levels: morphemes (in the seminal work by G. Altmann [16] that he cites) or sentences [18]; see Table 1. Probably the most radical example is music (see also Table 1), where the whole and the parts lack a "meaning" equivalent to that of content words. This suggests that Menzerath's law is a manifestation of abstract principles as many have proposed (see [12] and references therein; [14]). In contrast, Solé shows a lack of abstraction when considering that language and genomes, in order to resemble statistically, must be practically identical [10]. Indeed, he interprets the linguistic metaphor that inspired our original article (genomes "are" words and chromosomes "are" syllables) not as a metaphor but as a narrow equivalence. We could have replaced words and syllables by other units: morphemes and syllables, sentences and clauses, or mr-segments and F-motifs (Table 1). Words and syllables were probably the simplest metaphors for a general audience.

## 4. Discussion

We have seen that Menzerath's law is not inevitable in genomes and that if suffices that the number of parts (e.g., the number of chromosomes) and the size of the whole in the units of the parts (the size of chromosomes in bases) are correlated in order to reject a trivial case of the law [15,19]. However, we do not mean that the finding of a non-trivial Menzerath's law in the relationship between mean chromosome size and chromosome number [14,19] is due to the striking similarities between non-coding DNA and linguistic units with grammatical meaning that we have enlightened here but Solé neglected [10]. We have never argued that the finding of the law in genomes is indicative of meaning, syntax or any other important property of language. The finding of Menzerath's law both when non-coding DNA is excluded [15] and when non-coding and coding-DNA are mixed [14], and beyond, i.e. in language (see [33] for a review) and music [12], suggests that a higher level of abstraction is necessary for understanding the recurrence of the law.

To our knowledge, it has not been investigated yet if non-coding DNA alone could lead to Menzerath's law, or more interestingly, a non-trivial Menzerath's law. Without this research, it is not possible either to have a clearer understanding of the role of non-coding DNA in the emergence of Menzerath's law in genomes or to question the relevance of the law in genomes. Perhaps, rather than precluding the emergence of the law or leading to a trivial law, non-coding DNA may contribute to the emergence the law in a way that defies a trivial explanation.

Languages and genomes show a striking similarity at the semantic level: both possess units that have an arbitrary semantic reference of symbolic nature [38]. Our comparison goes further and suggests that genomes code for some abstract version of grammatical and lexical meaning, the former in non-coding regions and the latter in coding regions. However, the depth of the similarity and the possible DNA-specific properties must be investigated further. One of the challenges for language research is estimating the proportion of material with grammatical meaning including both free function words and bound morphemes.

Quantitative linguistics offers powerful tools for discovering and investigating non-trivial connections between human language and genomes [38,39]. However, the evolutionary mechanisms and the constraints that may underlie the recurrence of Menzerath's law still must be understood.

ACKNOWLEDGEMENTS

We are grateful to M. D. Jiménez-López, D. Searls, F. Bartumeus, D. Alonso and S. Caldeira for helpful discussions. This work was supported by the grant *Iniciació i reincorporació a la recerca* from the Universitat Politècnica de Catalunya and the grant BASMATI (TIN2011-27479-C04-03) from the Spanish Ministry of Science and Innovation (RFC and JB).

Table 1. Some parameters of Menzerath-Altmann law. The summary is based upon the pioneering work of G. Altmann and collaborators. *N. A.* means that the two parameter version of Eq. 1, with *c*=0, was fitted. (*) this follows the notation $\mu\pm\sigma$, where $\mu$ is the mean value of *b* in all samples and $\sigma$ is the corresponding standard deviation among samples.

| Type of source | Size of the whole (x) | Size of the parts (y) | Languages | Samples | b | c | Ref. |
|---|---|---|---|---|---|---|---|
| Language | Morpheme length (in syllables) | Mean syllable length (in phomenes) | Indonesian | 1 | -0.37 | 0.048 | [16] |
| | Word length (in syllables) | Mean syllable length (in phonemes) | English | 1 | 0.15 | -0.10 | [16] |
| | Sentence length (in clauses) | Mean clause length (in words) | German, English, French, Swedish, Hungarian, Slovak, Czech, Indonesian | 42 | -0.27 ± 0.11* | N.A. | [18] |
| Music | mr-segment length (in F-motifs) | Mean F-motif length (in tones) | - | 11 | -0.44 ± 0.09* | N.A. | [12] |

Table 2. The distance to $b$ = -1. A summary of |$b$+1|, the difference between the exponent $b$ obtained from the fit of $L_g \sim L_g^b$ and the exponent -1 that is expected from the arguments by Solé [10]. Groups are sorted increasingly by |$b$+1|. $b$ was estimated using non-linear regression as in [17]. The dataset is the same as that of [17,19]. The values of |$b$+1| were rounded to leave only two significant digits. (*) is used for the only two groups used by Solé [10]. Two interpretations of Sole's notion of plant are offered: angiosperms and a mixture of angiosperms and gymnosperms.

| Group | |$b$+1| |
|---|---|
| Mammals* | 0.014 |
| Birds | 0.042 |
| Angiosperm plants* | 0.051 |
| Plants* | 0.13 |
| Cartilaginous fishes | 0.18 |
| Reptiles | 0.39 |
| Insects | 0.31 |
| Jawless fishes | 0.45 |
| Ray-finned fishes | 0.46 |
| Fungi | 0.50 |
| Gymnosperm plants | 0.68 |
| Amphibians | 1.1 |

Table 3. Percentage of content, function and filler words in two registers: conversation and news report. Adapted from Table 2.4. of [27] (pp. 61).

|  | Conversation | News |
|---|---|---|
| Content words | 41% | 63% |
| Function words | 44% | 37% |
| Fillers | 15% | - |